\documentclass[10pt,letterpaper]{article}
\usepackage{opex3}
\usepackage{amssymb,amsmath}
\usepackage{color}
\usepackage{url}
\usepackage[all]{xy}
\usepackage{graphicx}
\usepackage{subfigure}
\usepackage{epsfig}
\usepackage{geometry}
\usepackage{listings}
\usepackage{booktabs}
\usepackage{footnote}
\usepackage{color}
\usepackage{booktabs}
\usepackage{array}
\usepackage[compress]{cite}
\newcommand{\Rom}[1]{\uppercase\expandafter{\romannumeral #1\relax}}
\begin{document}
\title{Optomechanical transductions in single and coupled wheel resonators} 

\author{Chenguang Huang$^*$, Jiahua Fan, Ruoyu Zhang, and Lin Zhu}

\address{The Holcombe Department of Electrical and Computer Engineering, 

Center for Optical Material Science and Engineering Technologies, 

Clemson University, Clemson, South Carolina, 29634, USA.}

\email{*chuang@clemson.edu}
\begin{abstract}
In this report, the optomechanical transductions in both single and two side-coupled wheel resonators are investigated. In the single resonator, the optomechanical transduction sensitivity is determined by the optical and mechanical quality factors of the resonator. In the coupled resonators, the optomechanical transduction is related to the energy distribution in the two resonators, which is strongly dependent on the input detuning. Compared to a single resonator, the coupled resonators can still provide very sensitive optomechanical transduction even if the optical and mechanical quality factors of one resonator are degraded.
\end{abstract}


\section{Introduction}
The coupling between the optical cavity modes and the mechancial modes via radiation pressure has been investigated in both fundamental and applied studies\cite{nearfield_optomechanics,accelerometer,cavity_optomechanics}. Optomechanical devices utilize the radiation pressure inherent in photons to sense and feed back mechanical motion across a wide range of experimental platforms\cite{wheel_design,beam_disk,beam_PhC,cantilever_array,cantilever_with_disk,study_of_parametric_instability}. The mostly studied optomechanical device of a single resonator usually involves the coupling between one optical mode and one mechanical mode, such as ring resonator\cite{Sid_OE_nitride_ring}, disk resonator\cite{disk_low_threshold}, microtoroid resonator\cite{toroid_oscillation} and zipper resonator\cite{optomechanical_crystal}. The coupling between one optical mode and few mechanical modes has also been studied in double ring resonators\cite{stacked_ring}, double disk resonators\cite{stacked_disk} and coupled zipper cavities\cite{coherent_mixing}. Although the composite optical resonance in coupled optical resonators is more complex and exhibits interesting physics, it has not been widely used for controlling optomechanical interactions\cite{synchronization}. Here we compare the optomechanical transductions in a single resonator and two side-coupled resonators, and show that the energy distribution in coupled optical resonators has a great impact on the optomechanical coupling. In addition, we show that the sensitive optomechanical transduction can still be obtained in coupled resonators despite the degraded optical and mechanical quality factors in one resonator.

\section{Device design and experimental setup}
\begin{figure}[tph]
\centering\includegraphics[scale=0.68]{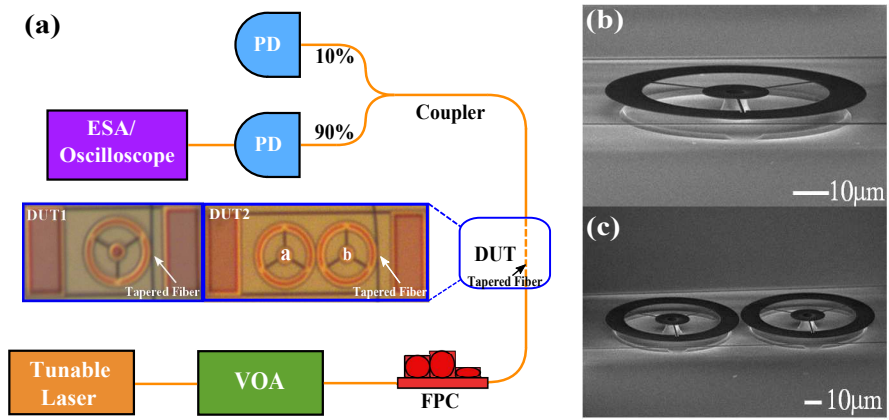}
\caption{(a) Experimental setup. VOA: Variable Optical Attenuator; FPC: Fiber Porlarization Controller; PD: Photodetector; ESA: Electrical Spectrum Analyzer; DUT: Device Under Test. A tapered fiber is used to couple light into and out of the resonator, and it is parked on the nanostrings. The experiments are conducted under vacuum. Both single wheel resonator(DUT1) and two side-coupled wheel resonators(DUT2) are investigated. (b) SEM image of DUT1. (c) SEM image of DUT2.}
\label{sem}
\end{figure}
We employ the same wheel resonator design as described in our earlier work\cite{rf_mixing_huang}. The $10\mu m$ wide wheel resonator has three thin spokes connecting to the center support and the large undercut ratio($90\%$) of the center support helps minimize the mechanical energy dissipation\cite{high_Q_undercut}. Figure 1(a) shows the experimental setup, where DUT1 and DUT2 represent the two different devices investigated in this report. DUT1 has a single wheel resonator, and DUT2 has two side-coupled wheel resonators(represented by $a$ and $b$ in Fig. 1(a)). Figure 1(b) and (c) show the SEM images of the two devices. The distance between the two coupled resonators is 600nm in Fig. 1(c). A tapered fiber is used for coupling light into and out of the resonators. All the experiments are conducted under vacuum. 
\section{Single resonator measurement results}
First, we test the single resonator(DUT1). Figure 2 shows the optical transmission spectra around 1571.22nm and the RF spectra of the fundamental breathing mode, respectively. The blue circle in Fig. 2(a) shows the optical transmission as the taper is placed around 400nm away from the resonator, indicating a high Q resonance with intrinsic Q of $4.8\times10^5$ and loaded Q of $4.3\times10^5$. Then we tune the input wavelength at appropriate detuning and keep the input power at $60\mu W$. The blue circle in Fig. 2(b) shows the RF spectrum of the fundamental breathing mode with mechanical quality factor of 4243. The sensitive optomechanical transduction is obtained at low input power due to the high loaded Q and weak mechanical damping induced by the large undercut ratio of the center support.
\begin{figure}[tph]
\centering\includegraphics[scale=0.7]{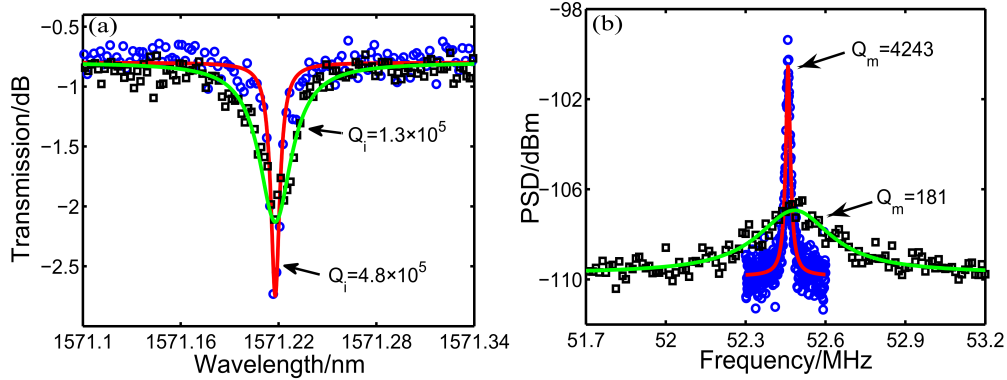}
\caption{Measurements for DUT1. (a) Cavity ransmision spectra with different taper couplings. The transmission spectrum denoted by the blue circle with red line fitting is obtained when the taper is 400nm away from the resonator. The transmission spectrum denoted by the black square with green line fitting is obtained when the taper touches the resonator. (b) RF spectra of the breathing mode when the taper is away from (blue circle) and in contact with the resonator (black square).}
\label{setup}
\end{figure}

When we place the taper in contact with the device, the optical transmission loss increases by around 3dB. The black square in Fig. 2(a) shows the optical transmission with intrinsic quality factor degraded to $1.3\times10^5$. The loaded quality factor is $1.2\times10^5$. Due to the degraded optical Q and increased mechanical damping induced by the taper, much higher input power is required to read out the RF signal on the spectrum analyzer. A different detector (New Focus 1611) is used for the RF signal extraction. The black square in Fig. 2(b) shows the RF spectrum of the breathing mode with mechanical quality factor of 181 at input power of $800\mu W$. The two RF singals in Fig. 2(b) are postprocessed to have the same level of noise background for comparison. The degradation of the optical quality factor and increased mechanical damping result in the less sensitive optomechanical transduction.
\section{Coupled resonator measurement results}
Then we test the two side-coupled resonators(DUT2). Figure 3(a) shows the optical transmission spectra from 1570nm to 1575nm when the taper seperately couples to the resonator $a$ and resonator $b$. The transmission spectrum colored in blue and red is taken when the taper couples to the resonator $a$. The transmission spectrum colored in black and green is taken when the taper couples to the resonator $b$. The resonance dips colored in red and green represent the shared coupled resonance. Due to fabrication errors, the two spectra also have different sets of uncoupled modes. The different extinctions at the coupled resonance are due to different coupling conditions.
\begin{figure}[tph]
\centering\includegraphics[scale=0.7]{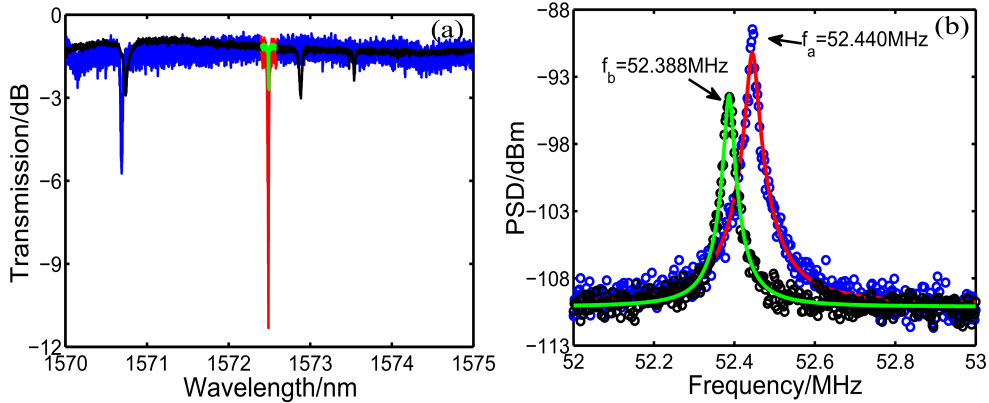}
\caption{Measurements for DUT2. (a) Cavity transmission spectra in a wide wavelength range. The transmission spectrum colored in blue and red is taken when the taper couples to the resonator $a$. The transmission spectrum colored in black and green is taken when the taper couples to the resonator $b$. The resonance dips colored in green and red represent the coupled resonance. (b) RF spectra at uncoupled resonances when the taper separately couples to resonator $a$ (blue circle) and resonator $b$ (black circle).}
\label{setup}
\end{figure}
At uncoupled resonances denoted by the black line in Fig. 3(a), only the optomechanical transduction in the resonator $b$ is detected as shown in the black circle in Fig. 3(b). $f_b$ represents the eigenfrequency of the breathing mode in the resonator $b$. When we probe the RF signal at uncoupled resonances denoted by the blue line in Fig. 3(a), only the optomechanical transduction in the resonator $a$ is detected as shown in the blue circle in Fig. 3(b). $f_a$ represents the eigenfrequency of the breathing mode in the resonator $a$. Due to fabrication errors, $f_a$ and $f_b$ are slightly different. For the rest of the measurements in this report, we mainly focus on the optomechanical transduction around the coupled resonance denoted by the green line at 1572.50nm shown in Fig. 3(a).
\begin{figure}[tph]
\centering\includegraphics[scale=0.68]{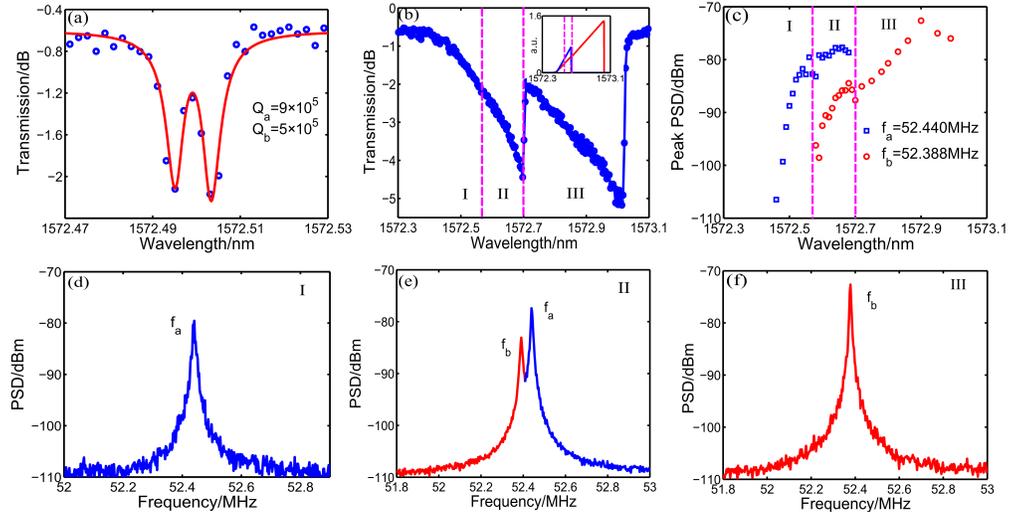}
\caption{Measurements for DUT2. (a) Experimentally recorded cavity transmission spectrum (blue circle) with theoretical fitting (red line) under input power of $4\mu W$. $Q_a=9\times10^5$, $Q_b=5\times10^5$. The taper is 400nm away from the device. (b) Cavity transmission with thermal effects under input power of $60\mu W$. \Rom{1},\Rom{2} and \Rom{3} represent three regions as the wavelength is swept across the resonances during the RF measurement. The subset shows the simulated stored energy distribution in the coupled resonators with aribitrary unit. The blue/red line represents the stored energy in resonator $a$/$b$. (c) PSD peak values of the RF spectrum for the breathing modes in the two resonators. $f_a$ and $f_b$ represent the frequencies of the breathing modes for the two resonators. (d-f) RF spectra at the input wavelength of 1572.56nm(d), 1572.66nm(e) and 1572.90nm(f).}
\label{setup}
\end{figure}

Figure 4(a) shows the detailed optical transmission spectrum at 1572.50nm. $Q_a$ and $Q_b$ represent the intrinsic quality factors of the two resonators. As shown in Fig. 4(a), the two resonance dips are slightly asymmetric due to the fact that the resonance wavelength in resonator $a$ is slightly shorter than that in resonator $b$. Figure 4(b) shows the optical transmission spectrum with thermal effects under input power of $60\mu W$. Since the taper is moved slightly closer to the resonator $b$, the extinction ratio is higher compared with Fig. 4(a). The subset in Fig. 4(b) shows the simulated energy distribution in the coupled resonators during the wavelength scanning\cite{thermal_coupled_resonators}. The blue/red line represents the stored energy in resonator $a$/$b$. RF signals are extracted while the input wavelength is swept across the resonances from 1572.48nm to 1572.95nm. The swept wavelength range is divided into three regions denoted by \Rom{1}, \Rom{2} and \Rom{3}. Figure 4(c) shows the recorded RF peak values while the wavelength is tuned from region \Rom{1} through region \Rom{3}. The optical power builds up in both resonators as we tune the wavelength upward from 1572.48nm. And the resonances of both resonators are red shifted due to thermal effects during the wavelength tuning. When the input wavelength is shorter than 1572.57nm(region \Rom{1}), the dropped optical power in resonator $a$ is higher than that in resonator $b$. Meanwhile, we could only detect the fundamental breathing mode of resonator $a$ in the RF spectrum. Figure 4(d) shows the RF spectrum at the input wavelength of 1572.56nm, indicating that only the optomechanical transduction in resonator $a$ is detected. 

When the input wavelength is between 1572.57nm and 1572.70nm(region \Rom{2}), the optical power in resonator $b$ is large enough to make the optomechanical coupling in resonator $b$ detectable in the ESA. Thus we could detect the fundamental modes of both resonator $a$ and $b$ in this wavelength range. Figure 4(e) shows the RF spectrum at the input wavelength of 1572.66nm, where both RF peaks at frequencies $f_a$ and $f_b$ are detected. In region \Rom{2}, the relative detuning in resonator $a$ is thermally locked due to the bistability, and the dropped power keeps almost the same in resonator $a$. The dropped power keeps increasing in resonator $b$. So the RF peak power at $f_a$ does not change much and the RF peak power at $f_b$ increases as the wavelength is tuned from 1572.57nm to 1572.70nm as shown in Fig. 4(c). The dropped power in resonator $a$ is always greater than that in resonator $b$ when the input wavelength is shorter than 1572.70nm. When the input wavelength is longer than 1572.70nm(region \Rom{3}), the heat power in the resonator $a$ is not enough to compensate the thermal dissipation, so its resonance returns to the cold cavity resonance and there is almost no optical energy build-up in resonator $a$. The optical power only builds up in resonator $b$. Figure 4(f) shows the RF spectrum at the input wavelength of 1572.90nm, and only one RF peak at $f_b$ is detected. In our experiments, the optical power reaching the detector is around 15$\mu W$ and the dropped power into the resonators ranges from 2$\mu W$ to 8$\mu W$. When the dropped power increases, both resonators experience weak optomechanical backaction effects\cite{optomechanical_crystal}. For the resonator $a$, the mechanical Q increases from 4700 to 5500 and the mechanical frequency decreases from 52.442MHz to 52.439MHz. For the resonator $b$, the mechanical Q increases from 4800 to 7000 and the mechanical frequency decreases from 52.388MHz to 52.374MHz.
\begin{figure}[tph]
\centering\includegraphics[scale=0.68]{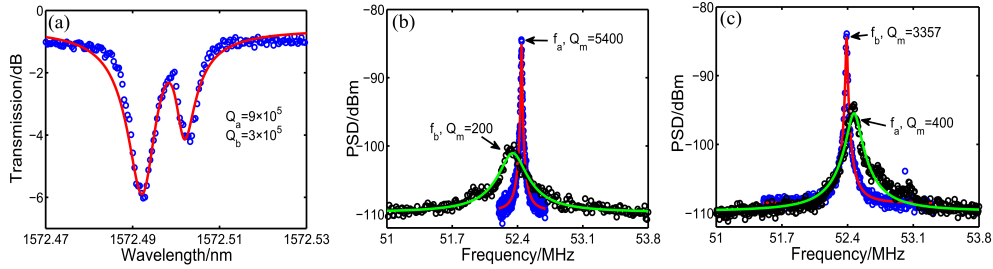}
\caption{Measurements for DUT2. (a) Cavity transmission spectrum when the taper is in contact with the resonator $b$. (b) RF spectra of the breathing modes when the taper is away from (blue circle) and in contact with the resonator $b$ (black circle). (c): RF spectra of the breathing modes when the taper is away from (blue circle) and in contact with the resonator $a$ (black circle).}
\label{setup}
\end{figure}

Then we place the taper in contact with the resonator $b$. Figure 5(a) shows the optical transmission spectrum, and the optical quality factor of the resonator $b$ degrades to $3\times10^5$ due to the taper induced optical loss. Figure 5(b) shows the RF spectra of the breathing modes of the two coupled resonators at input power of $60\mu W$, respectively. The mechanical quality factor of the resonator $b$ degrades by an order of magnitude due to the taper induced mechanical loss. Despite the degraded optical and mechanical quality factors, the breathing mode in resonator $b$ can still be sensitively detected at low input power in the ESA through the coupling to the undegraded optical resonance in resonator $a$. Figure 5(c) shows the RF spectra of the breathing modes of the two resonators when the taper is in contact with the resonator $a$ at input power of $60\mu W$. In contrast, the optomechanical transduction of the device DUT1 requires much higher input power when the taper touches the device as shown in Fig. 2(b). The different mechanical quality factors at the RF signals denoted by the blue circles in Fig. 5(b) and Fig. 5(c) are due to the fluctuation of the vacuum level during the measurement. Since it is difficult to control the exact position where the taper touches the resonator, the mechanical losses induced by the taper are different as shown by the different mechanical quality factors at the RF signals denoted by the black circles in Fig. 5(b) and Fig. 5(c).

\section{Conclusion}
In summary, we investigate the optomechanical transductions in both single and two side-coupled wheel resonators. In the single resonator, the degradation of mechanical and optical quality factors results in much less sensitive optomechanical transduction. In two side-coupled resonators, the energy distribution is modified by the input detuning and plays an important role in the optomechanical transductions. We also show that sensitive optomechanical transduction can still be obtained in coupled resonators even if the mechanical and optical quality factors are degraded in one resonator.


\section*{Acknowledgements}
The authors would like to thank Siddharth Tallur and Mian Zhang for the helpful discussions about the fabrication of the device. This work was performed in part at the Cornell Nanoscale Facility, a member of the National Nanotechnology Infrastructure Network, which is supported by the National Science Foundation.

\begin{thebibliography}{99}
\bibitem{nearfield_optomechanics}
G. Anetsberger, O. Arcizet, Q. P. Unterreithmeier, R. Riviere, A. Schliesser, E. M. Weig, J. P. Kotthaus and T. J. Kippenberg,"Near-field cavity optomechanics with nanomechanical oscillators,"
Nat. Phys.
\textbf{5},
909--914
(2009).

\bibitem{accelerometer}
A. G. Krause, M. Winger, T. D. Blasius, Q. Lin and O. Painter,"A high-resolution microchip optomechanical accelerometer,"
Nat. Photonics
\textbf{6},
768--772
(2012).

\bibitem{cavity_optomechanics}
T. J. Kippenberg and K. J. Vahala,"Cavity opto-mechanics,"
Opt. Express
\textbf{15},
17172--17205
(2007).

\bibitem{wheel_design}
X. Sun, K. Y. Fong, W. H. P. Pernice and H. X. Tang,"GHz optomechanical resonators with high mechanical Q factor in air,"
Opt. Express
\textbf{19},
22316--22321
(2011).

\bibitem{beam_disk}
O. Basarir, S. Bramhavar and K. L. Ekinci,"Monolithic integration of a nanomechanical resonator to an optical microdisk cavity,"
Opt. Express
\textbf{20},
4272--4279
(2012).

\bibitem{beam_PhC}
X. Sun, J. Zheng, M. Poot, C. W. Wong, and H. X. Tang,"Femtogram doubly clamped nanomechanical resonators embedded in a high-Q two-dimensional photonic crystal nanocavity,"
Nano Lett.
\textbf{12},
2299--2305
(2012).

\bibitem{cantilever_array}
O. Basarir, S.  Bramhavar and K. L. Ekinci,"Motion transduction in nanoelectromechanical systems (NEMS) arrays using near-field optomechanical coupling,"
Nano Lett.
\textbf{12},
534--539
(2012).

\bibitem{cantilever_with_disk}
K. Srinivasan, H. X. Miao, M. T. Rakher, M. Davanco and V. Aksyuk,"Optomechanical transduction of an integrated silicon cantilever probe using a microdisk resonator,"
Nano Lett.
\textbf{11},
791--797
(2011).

\bibitem{study_of_parametric_instability}
H. Rokhsari, T. J. Kippenberg, T. Carmon and K. J. Vahala,"Theoretical and experimental study of radiation pressure-induced mechanical oscillations(parametric instability) in optical microcavities,"
IEEE J. Sel. Top. Quantum Electron.
\textbf{12},
96--107
(2006).

\bibitem{Sid_OE_nitride_ring}
S. Tallur, S. Sridaran and S. A. Bhave,"A monolithic radiation-pressure driven, low phase noise silicon nitride opto-mechanical oscillator,"
Opt. Express
\textbf{19},
24522--24529
(2011).
%

\bibitem{disk_low_threshold}
W. C. Jiang, X. Lu, J. Zhang, and Q. Lin,"High-frequency silicon optomechanical oscillator with an ultralow threshold,"
Opt. Express
\textbf{20},
15991--15996
(2012).
%

\bibitem{toroid_oscillation}
T. J. Kippenberg, H. Rokhsari, T. Carmon, A. Scherer and K. J. Vahala,"Analysis of radiation pressure induced mechanical oscillation of an optical microcavity,"
Phys. Rev. Lett.
\textbf{95},
033901
(2005).
%

\bibitem{optomechanical_crystal}
M. Eichenfield, J. Chan, R. M. Camacho, K. J. Vahala and O. Painter,"Optomechanical crystal,"
Nature
\textbf{462},
78--82
(2009).
%

\bibitem{stacked_ring}
G. S. Wiederhecker, L. Chen, A. Gondarenko and M. Lipson,"Controlling photonic structures using optical forces,"
Nature
\textbf{462},
633--636
(2009).
%

\bibitem{stacked_disk}
Q. Lin, J. Rosenberg, X. Jiang, K. J. Vahala and O. Painter,"Mechanical oscillation and cooling actuated by the optical gradient force,"
Phys. Rev. Lett.
\textbf{103},
103601
(2009).
%

\bibitem{coherent_mixing}
Q. Lin, J. Rosenberg, D. Chang, R. Camacho, M. Eichenfield, K. J. Vahala and O. Painter,"Coherent mixing of mechanical excitations in nano-optomechanical structures,"
Nat. Photonics
\textbf{4},
236--242
(2010).
%

\bibitem{synchronization}
M. Zhang, G. S. Wiederhecker, S. Manipatruni, A. Barnard, P. McEuen, and M. Lipson,"Synchronization of micromechanical oscillators using light,"
Phys. Rev. Lett.
\textbf{109},
233906
(2012).
%

\bibitem{rf_mixing_huang}
C. Huang, J. Fan, R. Zhang, and L. Zhu,"Internal frequency mixing in a single optomechanical resonator,"
Appl. Phys. Lett.
\textbf{101},
231112
(2012).
%

\bibitem{high_Q_undercut}
X. Sun, X. Zhang and H. X. Tang,"High-Q silicon optomechanical microdisk resonators as gigahertz frequencies,"
Appl. Phys. Lett.
\textbf{100},
173116
(2012).
%

\bibitem{thermal_coupled_resonators}
C. Huang, J. Fan and L. Zhu,"Dynamic nonlinear thermal optical effects in coupled ring resonators,"
AIP Advances
\textbf{2},
032131
(2012).
%
\end{thebibliography}
\end{document}